\documentclass[aps,prb,twocolumn,nobibnotes]{revtex4-1}  
\usepackage{dcolumn}
\usepackage{graphicx}
\usepackage{color}
\usepackage{amssymb}
\usepackage{amsmath}
\usepackage{braket}
\usepackage{bm}
\usepackage{soul}
\usepackage[colorlinks=true,linktocpage=true,breaklinks=true]{hyperref}

\DeclareUnicodeCharacter{2212}{-}

\usepackage{siunitx}
\usepackage{textcomp}

\begin{document}

\title{Spin and Orbital Rashba response in ferroelectric polarized PtSe$_2$/MoSe$_2$/LiNbO$_3$ heterostructures}


\author{A.~Pezo}
\affiliation{Laboratoire Albert Fert, CNRS, Thales, Université Paris-Saclay, 91767 Palaiseau, France}
\author{J. Li}
\affiliation{Univ. Grenoble Alpes, CEA, Leti, F-38000 Grenoble, France}
\author{F. Ibrahim }
\author{M. Jamet}
\author{M.~Chshiev}
\affiliation{Univ.~Grenoble Alpes, CEA, CNRS, Grenoble INP, IRIG-SPINTEC, F-38000 Grenoble, France}
\author{S.~Massabeau}
\author{F.~Miljevic}
\author{J.-M.~George}
\author{H.~Jaffrès\footnote{corresponding author: henri.jaffres@cnrs-thales.fr}}
\affiliation{Laboratoire Albert Fert, CNRS, Thales, Université Paris-Saclay, 91767 Palaiseau, France}

\begin{abstract}

Recent studies using Terahertz Time-Domain Spectroscopy (THz-TDS) with spintronic emitters as a source have revealed distinct signatures of the Rashba effect. This effect, which arises from the breaking of inversion symmetry in low-dimensional materials, has been recently investigated in CoFeB/PtSe$_2$/MoSe$_2$/LiNbO$_3$-based heterostructures [\textit{S. Massabeau et al., APL Mater. 13, 041102, 2025}]. The observed phenomena are at the source of the generated THz far-field emission, typically through mechanisms such as spin-to-charge conversion triggered by the absorption of ultrafast optical pulses. In this work, we employ first-principles simulations to quantify the Rashba effect at PtSe$_2$/MoSe$_2$/LiNbO$_3$ interfaces, expanding the traditional understanding of spin transport by incorporating the orbital degree of freedom. Moreover, we quantify the degree of control on the THz emission depending on the polarization direction of LiNbO$_3$. In order to achieve this, we analyze the accumulation of both spin and orbital components using linear response theory, revealing distinct behaviors. These findings are crucial for a deeper understanding of the physical processes governing angular momentum-to-charge conversion and THz emission. Moreover, they may provide broader insights into various experimental outcomes, including those related to spin-orbit torque.

\end{abstract}

\maketitle

\section{Introduction}

Recent important advances in material growth and experimental techniques have allowed one to probe and exploit the dynamics of the spin degree of freedom, in particular for terahertz (THz) generation. These efforts lead to a new class of terahertz sources based on ultrafast spin-to-charge conversion (SCC) effects, commonly referred to as spintronic THz emitters~\cite{Seifert2017,Rongione2023,Yadav2024,Massabeau2025}. Since the THz spectral region is important for future high-speed telecommunications, including applications in wireless communication, nano-networks, on-chip interconnects, and advanced THz spectroscopy and imaging~\cite{THz_9508905,Sitnikov}, the development of future devices would guide future investigations in materials science. In this sense, spintronics THz emitters typically consist of nanometer-thick heterostructures made up of a ferromagnetic (FM) layer interfaced with a non-magnetic (NM) layer~\cite{Zhou2018}. Owing the absorption of an ultrashort optical pulse by the heterostructure, a non-equilibrium spin current is generated within the FM layer, which then diffuses into the NM layer where it is converted into a transient charge current via a spin-to-charge conversion mechanism on the NM side or at the FM/NM interface, resulting in the emission of a linearly polarized THz pulse~\cite{Seifert2016,Jungfleisch2018,Dang2020,chia2021}. Currently, it is assumed that there exist two main phenomena responsible for SCC within THz emitters, one of them being the inverse spin Hall effect (ISHE) taking place whenever the NM displays strong spin-orbit coupling (\textit{e.g.}, Pt, Ta, W) or the inverse Rashba–Edelstein effect (IREE) at FM/NM interfaces ~\cite{Manchon2019}. The latter effect is especially involved in systems with spin-polarized surface states such as topological insulators as shown both experimentally~\cite{wang2018,Rongione2022,Rongione2023,bierhance2025} and theoretically~\cite{PhysRevResearch.6.043332,Anadon2025} and graphene-based heterostructures~\cite{Anadon2025,Abdukayumov_2025}. For this reason, extensive research has focused on improving the conversion process and thus THz emission efficiency, by optimizing parameters such as material selection, layer thicknesses, and the design of multilayered stacks. These advancements have led to emitters with output powers that surpass those of conventional THz sources like nonlinear optical crystals \cite{Rouzegar_2025}, positioning spintronic emitters as strong contenders for next-generation THz technologies. However, in order to fully develop their potential, it is essential to achieve dynamic control over THz emission using external degrees of freedom, therefore enabling information encoding. Despite significant advancements in this field, only a limited number of studies have tackled this challenge~\cite{Ji2023,Nkeck2022}, with one of the main approaches being emitter patterning and complex optical excitation schemes. However, these methods often come with inherent drawbacks, such as introducing permanent structural constraints or necessitating bulky, energy-intensive setups limiting the scalability and flexibility of spintronic THz devices. In such a scenario, considering solely ultrathin van der Waals(vdW) heterostructures and acting on the stacking/layers number would enable the possibility to tune and tailor emerging properties from 2D materials~\cite{Abdukayumov2024}. In particular, it has been shown that the spin-to-charge conversion properties – and thus THz emission – in MoSe2/PtSe2 vdW heterostructures on a ferroelectric substrate made of LiNbO3 have been shown to be modulated by the polarization direction~\cite{Massabeau2025}. 
At the same time, the generation and injection of spin currents is mandatory for spintronic devices, as those may efficiently transmit spin information in nonlocal magnetic spin-valves and convey spin torque in non-volatile memories (MRAM technologies) and nano-oscillators~\cite{Bhatti2017}. On the electronic and spin transport side, the two main mechanisms to generate pure spin currents include the spin Hall effect (SHE)  and the adiabatic spin pumping. The former occurs when a charge current induces a transverse flow of spin angular momentum~\cite{Sinova2015}, while its reciprocal effect, the ISHE, is captured by spin pumping measurements~\cite{Rojas-Sanchez2014}, referring to the generation of a non-equilibrium polarized current triggered by periodically varying a parameter in a conducting system. This setup has been developed in a bilayers composed of a ferromagnet and a nonmagnetic metal~\cite{Tserkovnyak2002,Brataas2002}. Consequently, this phenomenon has been investigated across various magnetic interfaces including transition metal ferromagnets, ferrimagnetic insulators, or antiferromagnets~\cite{PhysRevB.93.224415,PhysRevLett.113.057601}. Additionally, we may also exploit the Rashba-Edelstein effect (REE) at FM/NM interfaces to transform spin-to-charge currents. Interestingly, recent developments have introduced alternatives for generating pure spin-polarized currents, emphasizing the importance of the orbital degree of freedom. In recent years, theoretical and experimental demonstrations of the Orbital Hall effect~\cite{Go2018,Pezo2022,Salemi2022} have prompted the spintronic community to consider orbital pumping as a method to generate an orbital angular momentum~\cite{go_pumping,pezo_pumping}. This effect was experimentally observed in THz emission measurements in Ti/Fe and Ti/Ni bilayers~\cite{Choi2023}, wherein, given the negligible value of spin-orbit coupling (SOC) in the atomic elements, it was suggested the possibility to pump angular orbital momentum from the dynamics of magnetization~\cite{Hayashi2024}. By accounting for such contribution, the orbital counterpart to the IREE, the inverse orbital Rashba-Edelstein effect(IoREE) enters into play, whereby an orbital current is transformed into a charge flow~\cite{Go2021,Krishnia2024,Nikolaev2024,pezo2025anatomytorquesorbitalrashba}, moreover, recent studies have pointed out the implications of the orbital Rashba effect in HfO$_2$/Ni interfaces \cite{PhysRevB.111.L140405}. It is yet important to note the qualitative differences between the two degrees of freedom. Pure spin currents face inherent challenges because of their high sensitivity to relaxation processes caused by spin-orbit coupling and magnetic disorder. In contrast, the absence of direct coupling between orbital angular momentum and exchange magnetization provides a significant opportunity to convert orbit-to-charge currents more efficiently ~\cite{PhysRevResearch.2.033401,PhysRevB.108.075427}. 
Thus, a proper understanding of orbital dynamics in materials is crucial for developing new technologies based on this concept.

\vspace{0.1in}

This work is organized as follows: First, we introduce the methodology used to study the structural and out-of-equilibrium properties of these heterostructures by placing them in contact with a ferroelectric LiNbO$_3$ slab. This setup allows us to investigate the role of the polarization field within the heterostructure. From this analysis, we extract the effective out-of-plane electric field generated by the ferroelectric substrate, which we discuss in the next section. Next, we explore how this extracted value can be applied as an electric field in a pure MoSe$_2$/PtSe$_2$ bilayer. Following this, we present the necessary post-processing calculations to determine the spin and orbital accumulations in this context and conclude with our perspectives and insights on the findings.


\section{Methodology}

For studying the electronic properties of the considered multilayers and interfaces, we employed fully relativistic density functional theory (DFT). The structural relaxation and effective electric field extraction were carried out using DFT-PBE in plane-wave basis set with cut-off energy $500$~eV, Gamma point sampling in Brillouin zone, and D2 van der Waals correction, as implemented in VASP~\cite{Kresse_1993, Kresse_1996a, Kresse_1996b, Kresse_1999}. In order to calculate the transport properties, we describe the spin-orbit coupling within a fully relativistic pseudo-potential formulation and use the generalized gradient approximation for the exchange-correlation functional. The calculations are converged for a 400~Ry plane-wave cutoff for the real-space grid with a $(13 \times 13 \times 1)$  $\vec{k}$-points sampling of the Brillouin zone. On the other hand, we take advantage of the localized basis set as defined within SIESTA ~\cite{siesta_method} to study the transport properties. Similarly, we have employed the Perdew-Burke-Ernzerhof ~\cite{pbe,gga} exchange-correlation functional. The out-of-equilibrium transport calculations use a full \textit{ab-initio} DFT Hamiltonian matrices obtained directly from \textsc{SIESTA}~\cite{siesta_method} conveniently exploiting  the atom-centered double-$\zeta$ plus polarization (DZP) basis sets allowing its further manipulation for its usage with the Kubo formalism. We use the energy cutoff for real-space mesh of \SI{400}{Ry}. The self-consistent SOC is introduced via the full off-site approximation~\cite{Cuadrado_2012} using fully-relativistic norm-conserving pseudopotentials ~\cite{tm_pseudopotentials}. The system Hamiltonian and overlap matrices are obtained after performing a full self-consistent cycle and treated after within a post-processing routine utilizing SISL as an interface tool ~\cite{zerothi_sisl}.

\section{RESULTS AND DISCUSSIONS}

\subsection{Extraction of effective electric field in LiNbO$_3$/PtSe$_2$/MoSe$_2$}

\begin{figure*}[!ht]
    \centering    \includegraphics[width=1.\linewidth]{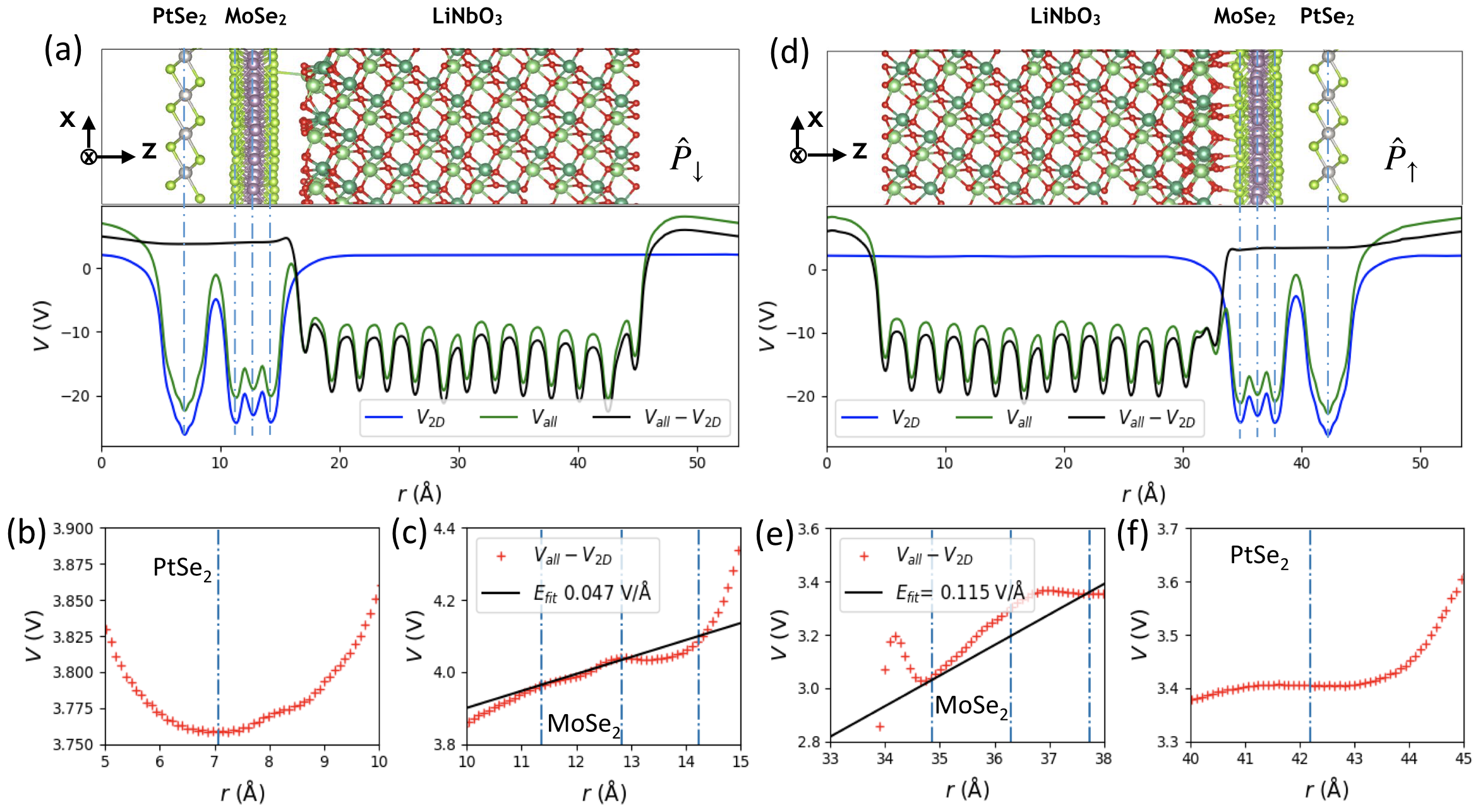}
    \caption{Extraction of the effective out-of-plane electric field from ferroelectric LiNbO$_3$ layers. In (a,d) we show the relaxed heterostructure with MoSe$_2$ and PtSe$_2$ layers below (a) and above (d) LiNbO$_3$ while keeping the same convention for the axes allowing us to define the polarizations $\hat{P}_{\uparrow,\downarrow}$ directions, therefore, the electric field used to simulate the polarization without LiNbO$_3$ should have also an opposite direction by accounting for the stacking of layers in (d). The in-plane averaged electrostatic potentials are plotted for the entire heterostructure (green line), and 2D layers only (blue line). The difference between the two potentials (black line) is the potential contributed solely by the ferroelectric LiNbO$_3$. (b) and (c) are the zoom-in of the potential difference for PtSe$_2$ and MoSe$_2$. The effective electric field is the gradient extracted in the MoSe$_2$ layer. The results of the opposite polarization (with 2D layers above LiNbO$_3$) are shown in (d) and with zoom-in potential in (e) and (f) for MoSe$_2$ and PtSe$_2$, respectively. in this sense, we emphasize that the electric field considered in (e) should have an opposite sign while considering the stacking in (a).}
    \label{fig:field}
\end{figure*}

The PtSe$_2$/MoSe$_2$/LiNbO$_3$ heterostructure is composed of a thick LiNbO$_3$ slab designed to retain its polarization field. The LiNbO3 slab arrangement represents in our case a Z-cut LiNbO3 crystal, with an out-of-plane ferroelectric polarization, normal to the LiNbO3 surface. Such a structure is made from a $3 \times 3$ in-plane supercell and has an approximate thickness of 3 nm, including a monolayer of MoSe$_2$ with in-plane lattice vectors of $(5, -1)$ and $(1, 4)$, as well as a monolayer of a $4 \times 4$ PtSe$_2$ supercell. A vacuum of 1.5 nm has been inserted into the unit cell, while the in-plane lattice parameter is fixed by LiNbO$_3$ at about 15.63~\AA. 
To quantify the effective electric field in the 2D layers contributed by the ferroelectric LiNbO$_3$ "substrate" at two different polarization states, the 2D layers are placed either above or below the LiNbO$_3$ slab. During the relaxation process of the two heterostructures, the atoms in the four layers of LiNbO$_3$ that are farthest from MoSe$_2$ are held fixed to simulate bulk LiNbO$_3$. The remaining atoms are relaxed until the force on them is below 10 meV/~\AA.




Figure~\ref{fig:field}(a) displays the relaxed structure of the PtSe$_2$/MoSe$_2$ layers positioned beneath LiNbO$_3$. The corresponding averaged in-plane electrostatic potentials are plotted along the out-of-plane direction (defined as the $z$ direction in this study). The potential of the entire structure, represented by the green line, exhibits dips at the positions of the atomic planes, as marked by the vertical dashed blue lines for the 2D layers.\\

To isolate the electric field contributed by the LiNbO$_3$ ferroelectric layer at the MoSe$_2$/PtSe$_2$ interface, we subtracted the potential of an isolated PtSe$_2$/MoSe$_2$ structure (depicted by the blue line) from that of the entire heterostructure. The resulting potential difference is shown by the black line and is zoomed in on the PtSe$_2$ region in Figure~\ref{fig:field}(b). Here, it appears almost constant, while for MoSe$_2$, as illustrated in Figure~\ref{fig:field}(c), the electric field is approximately $0.047$ V/\AA.


\vspace{0.1in}

The relaxed heterostructure with the opposite arrangement, specifically LiNbO$_3$/MoSe$_2$/PtSe$_2$, is illustrated in Fig.~\ref{fig:field}(d). This configuration is equivalent to reverse the polarization in LiNbO$_3$. The potential difference across the entire heterostructure, compared to the structure consisting solely of MoSe$_2$/PtSe$_2$, is displayed in Fig.~\ref{fig:field}(f) for PtSe$_2$. The potential difference is nearly constant, similar to the previous case. Additionally, for MoSe$_2$, it demonstrates an average electric field of $0.115$~V/~\AA, as shown in Fig.~\ref{fig:field}(e). This is a result of the shorter distance between MoSe$_2$ and LiNbO$_3$ when we consider opposite surfaces in the ferroelectric material, leading to $1.74$~\AA, compared to $2.61$~\AA, as found in the previous scenario.


\subsection{Spin and Orbital Rashba textures and Rashba-Edelstein linear responses}

With the knowledge of the effective electric fields arising from the ferroelectric material, we have performed \textit{ab-initio} simulations based on a localized basis for calculating the out-of-equilibrium properties such that we consider a trilayer system made out of two layers of PtSe$_2$ and a single MoSe$_2$ layer, where an artificial electric field was included during the self-consistent calculation to emulate the effect of the ferroelectric polarization field. In general, SCC occurring at interfaces may be more accurately described by the REE, where breaking inversion symmetry can lead to the locking of angular momentum (either spin or orbital). On the other hand, the spin Hall effect and REE are typically assigned to different origins in terms of \textit{interband} vs. \textit{intraband} quantum transitions. These phenomena are complementary in various contexts, such as spin-orbit torques (SOT) and their reciprocals effects manifesting as spin or/and orbital pumping in both ferromagnetic resonance (FMR) and THz emission regimes. The Inverse spin Rashba-Edelstein effect (iSREE) and its modulation by the ferroelectric polarization have already been addressed experimentally in our previous work~\cite{Massabeau2025}; nevertheless as recently emphasized~\cite{rappoport2021,roche2025}, the inverse orbital REE (iOREE) response for 2D materials have been overlooked in the literature and will be addressed in the following analysis.

\begin{figure*}[!ht]
    \centering    \includegraphics[width=1.\linewidth]{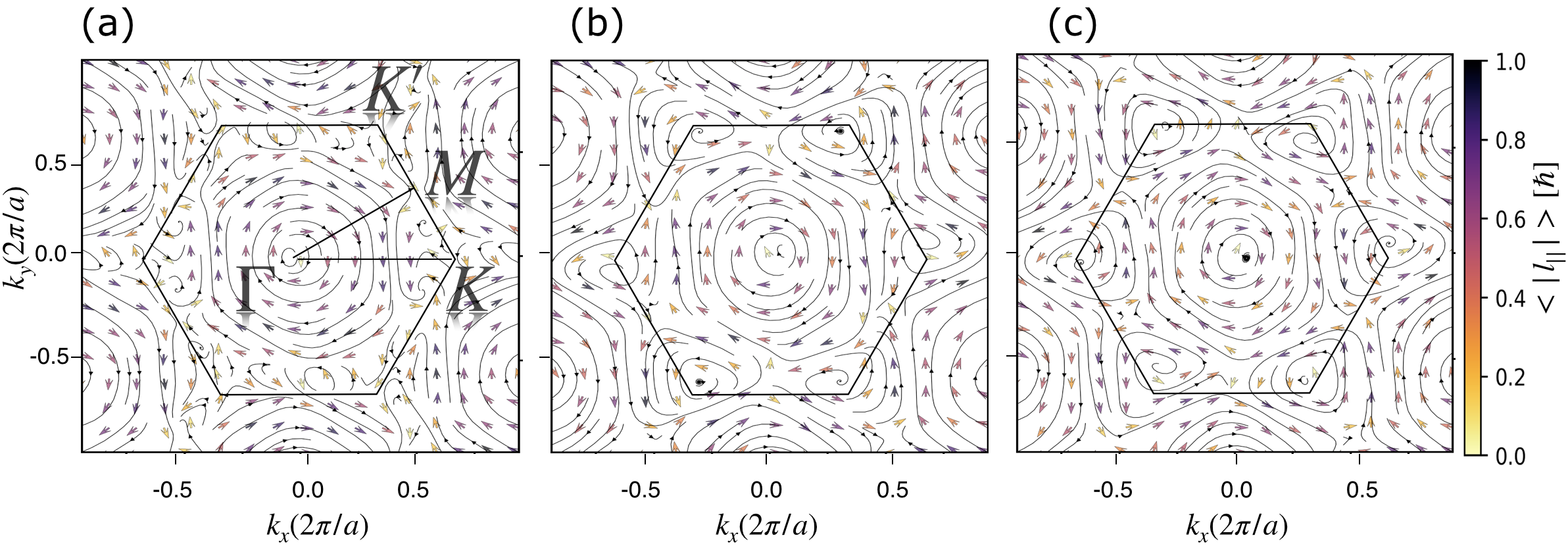}
    \caption{PtSe$_2$/MoSe$_2$ Orbital textures calculated at $E_F=-1.0$ eV for the three cases considered in this study, that is for the case of having the polarization field $\hat{P}_{\downarrow}$ (a), $\hat{P}=0$ (b) and $\hat{P}_{\uparrow}$ (c).}
    \label{fig:O_texture}
\end{figure*}

\begin{figure*}[!ht]
    \centering    \includegraphics[width=1.\linewidth]{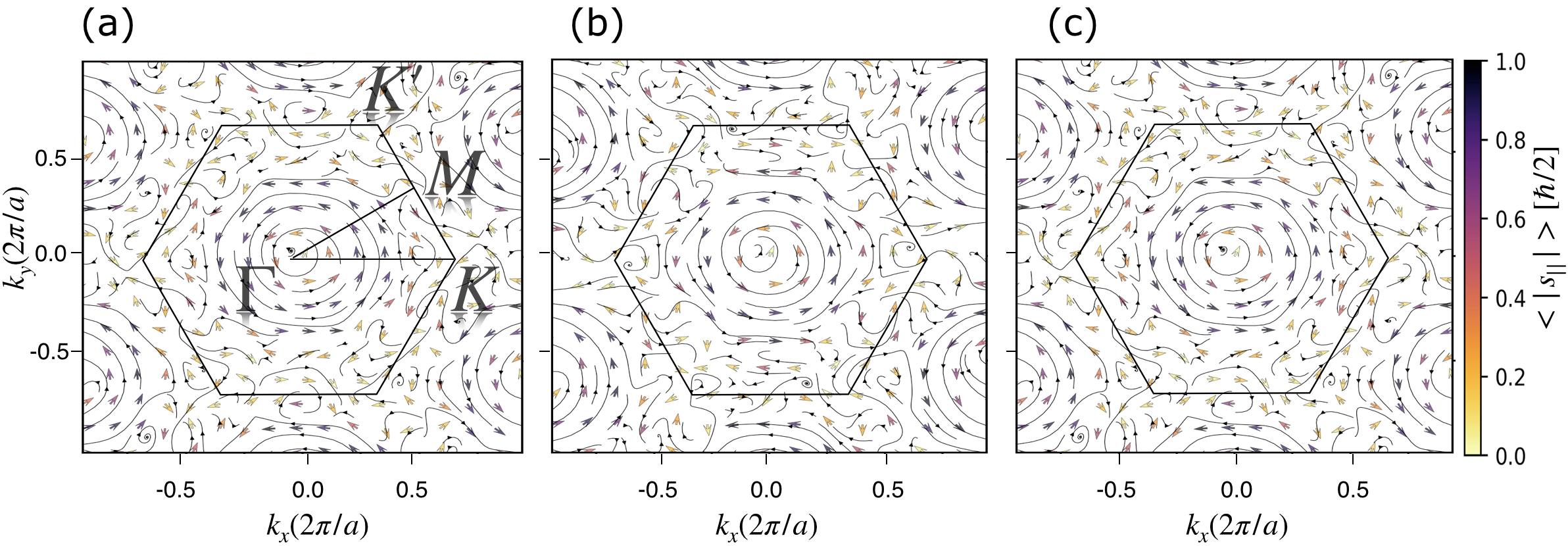}
    \caption{PtSe$_2$/MoSe$_2$ Spin textures calculated at $E_F=-1.0$ eV for the three same three cases depicted previously, that is, the polarization field $\hat{P}_{\downarrow}$ (a), $\hat{P}=0$ (b) and $\hat{P}_{\uparrow}$ (c).}
    \label{fig:S_texture}
\end{figure*}

\vspace{0.1in}

Regarding the SCC taking place in THz-TDS experiments using spintronics emitters as a source, the calculation of the inverse Rashba-Edelstein tensor, the reciprocal of the electrical driven effect, is not straightforward within the framework of linear response theory. This is because the external excitation is not an extensive quantity, as is the electric field $\mathcal{E}$ is needed for the calculation of the direct tensor~\cite{Johansson2021,PhysRevResearch.6.043332}. Instead, it is an intensive quantity, that is, the unbalanced angular momentum density (spin or orbital) carried by the excited non-equilibrium carriers. In the past, this was generalized to the expression for the REE length~\cite{Fert2013,Varotto2022,Johansson2021} derived from a refined linear response theory~\cite{Rongione2023}. This methodology evaluates both charge current and angular momentum density responses to an external time-varying magnetic field $\mathbf{k}$-dependent pumping excitation ~\cite{shen2014,PhysRevResearch.6.043332}. The advantage of this approach is that any anisotropy of electron scattering on the Fermi surface is related to both spin and orbital momentum locking is automatically incorporated. However, as such equivalence was previously demonstrated~\cite{Anadon2025}, we focus on the generated spin and orbital densities out-of-equilibrium once an electric field $\mathcal{E}$ is applied. As noted in a previous work~\cite{Abdukayumov2024}, even when holes show a reduced velocity in transition metal, the THz emission is likely to be originated from the emergence of the SCC within the valence band. However, the exact amount of spin polarized carriers injected and transmitted from the ferromagnetic side remains an unknown quantity, as this should depend on interfacial details such as roughness, crystalline and chemical defects, etc \cite{Rongione2023,pezo_pumping}.

In order to access such properties, we can examine the impact of external ferroelectric polarization on the vdW heterostructure for two different scenarios considering the mean value of the orbital and spin textures defined as $\braket{\psi_{\mathbf{k}n}|\hat{O}|\psi_{\mathbf{k}n}}$ where $\hat{O}$ represents either of both operators. First, as illustrated in Fig.~\ref{fig:O_texture}(b), the orbital texture varies depending on the polarization field: in case (a), $\hat{P}$ along $-\hat{z}$ (we note $\hat{P}_{\downarrow}$), while in case (c), $\hat{P}$ is along $+\hat{z}$ (we note $\hat{P}_{\uparrow}$). The case in (b) represents the absence of any polarization, since we have initially considered a trilayer system where the electric field was turned off
within the DFT calculations. Notably, the chirality near the $\Gamma$ point remains relatively unchanged, whereas there is noticeable distortion near the $K$ and $K'$ symmetry points. This distortion is especially pronounced in case (a), which exhibits a more pronounced three-fold symmetry compared to the picture in case (c). Moreover, when comparing Fig.~\ref{fig:O_texture} with its spin counterpart displayed in Fig.~\ref{fig:S_texture}, we observe that the orbital texture takes on a more hexagonal shape, with quite similar chiralities near the $\Gamma$ point when comparing spin and orbit. However, alike the spin, the chirality of the orbital texture at the $K$ and $K'$ points are counter-clockwise while $\Gamma$ displays a clockwise sense. As we will see below, this agrees with the reduction in the orbital polarization (accumulation) within the considered energy window, which results from the compensation of textures at high-symmetry points $\Gamma$ and $K(K')$ within the Brillouin zone, while the outcome in the case of spin is additive leading to a larger effect. Following such analysis, using a previously tested routine, we were able to calculate both spin and orbital polarization densities (accumulations)
in \ref{subsection:transport_PtSe2}. Under the application of an electric field along the $\hat{x}$ direction, it results a spin and orbital accumulations along $\hat{y}$. The current-driven spin and orbital densities are computed within the linear response formalism considering the symmetrized decomposition of Kubo formula \cite{Bonbien2020}, having the form:

\begin{equation*}
   \braket{\delta \hat{O}_y}=-e \mathcal{E}_x \tau  \int_{\mathbf{k}n} \partial_\epsilon f(\epsilon)\braket{\psi_{\mathbf{k}n}|\hat{v}_x|\psi_{n\mathbf{k}}} \braket{\psi_{\mathbf{k}n}|\hat{O}_y|\psi_{\mathbf{k}n}}
\end{equation*}

\noindent with $\hat{O}=\hat{s}$ or $\hat{O}=\hat{l}$ the respective spin and orbital operators, $v_x=(1/\hbar) \partial_{k_x} \hat{\mathcal{H}}$ the velocity operator, $f(\epsilon)$ is the equilibrium Fermi distribution function and $\ket{\psi_{n\mathbf{k}}}$ an eigenstate of the Hamiltonian $\hat{\mathcal{H}}$. The Brillouin zone integration was carried out on a $200 \times 200$ $\vec{k}-$meshgrid.

\vspace{0.1in}

\subsubsection{Rashba-Edelstein effect (REE) in PtSe$_2$/Gr as the reference heterostructures}

\begin{figure}[h]
    \centering    \includegraphics[width=1\linewidth]{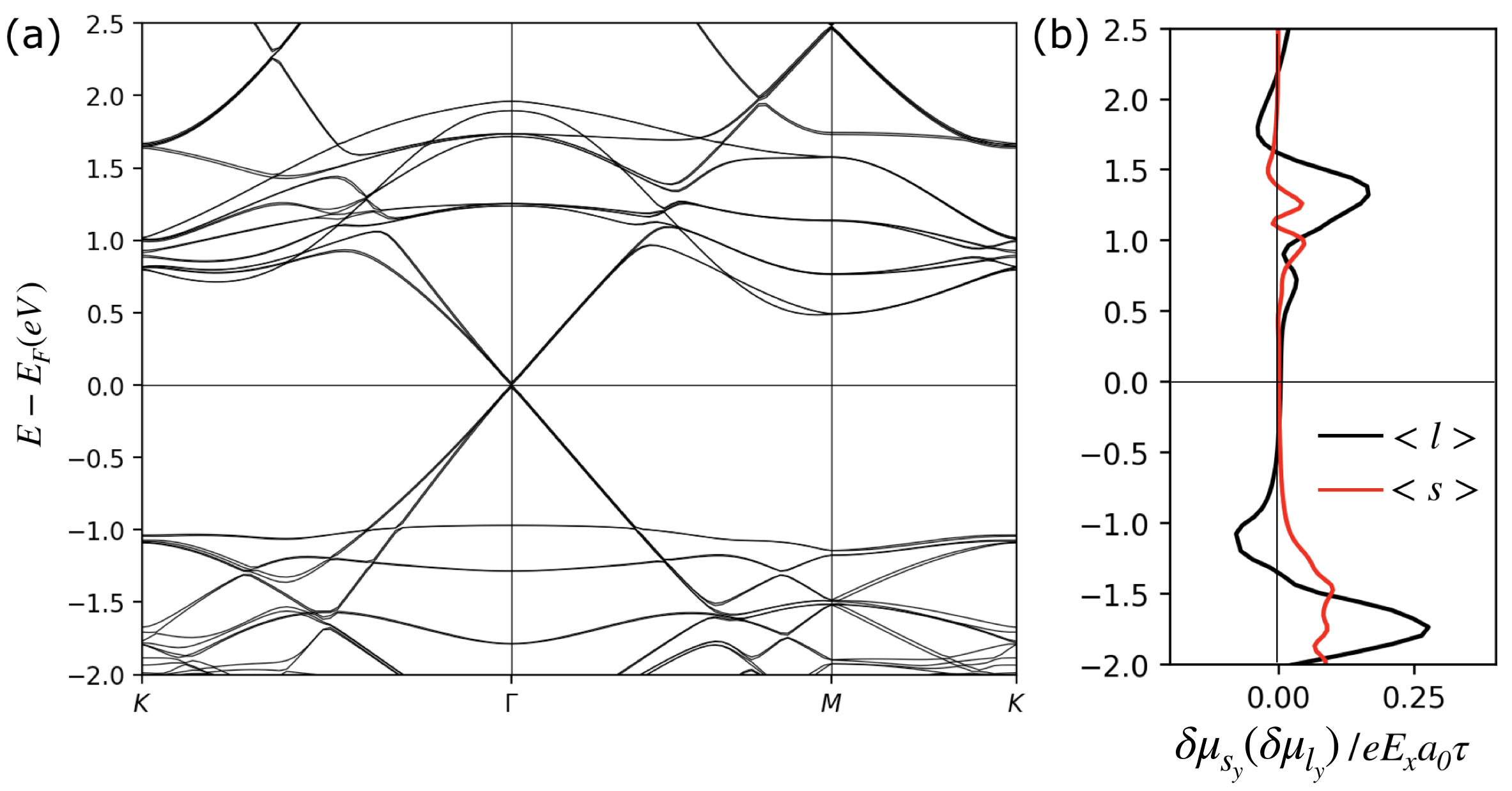}
    \caption{(a) Electronic band structure of PtSe$_2$/Graphene along the symmetry path $K-\Gamma-M-K$, (b) Out-of-equilibrium spin and orbital polarization densities (accumulation) using linear response for Gr/PtSe$_2$. In this case, one can observe the positive value obtained for both responses around the same energy window $[-1.5,0]$ as found experimentally in a previous report \cite{Abdukayumov2024}.}
    \label{fig:ptse2_Gra}
\end{figure}

Prior to considering the case of the interface between PtSe$_2$ and MoSe$_2$, we benchmark our methodology by showing our calculations performed for the case of PtSe$_2$/Graphene. The results are depicted in Fig.~\ref{fig:ptse2_Gra}, where the electronic band structure along specific lines of the 2D Brillouin zone is displayed in Fig.~\ref{fig:ptse2_Gra}(a) by side to the calculation of the spin and orbital polarization densities (accumulations) in Fig.~\ref{fig:ptse2_Gra}(b). In both cases, we have considered an energy window between -2~eV and +2~eV.  which encompasses the incident optical pulse energy of 1.55~eV used in a recent experimental report\cite{Abdukayumov2024} that is triggering the spin-to-charge conversion and THz emission.
Importantly, one has to emphasize the agreement between the sign of the spin and orbital accumulations with the THz signal reported previously as a result of the spin-to-charge conversion~\cite{Abdukayumov2024}. As described below, however, we should also consider the role of spin and orbital injections arising from the FM, this is clarify in the next section.

\subsubsection{Rashba Edelstein effect in PtSe$_2$/MoSe$_2$ heterostructures}\label{subsection:transport_PtSe2}

\begin{figure}[!ht]
    \centering    \includegraphics[width=\linewidth]{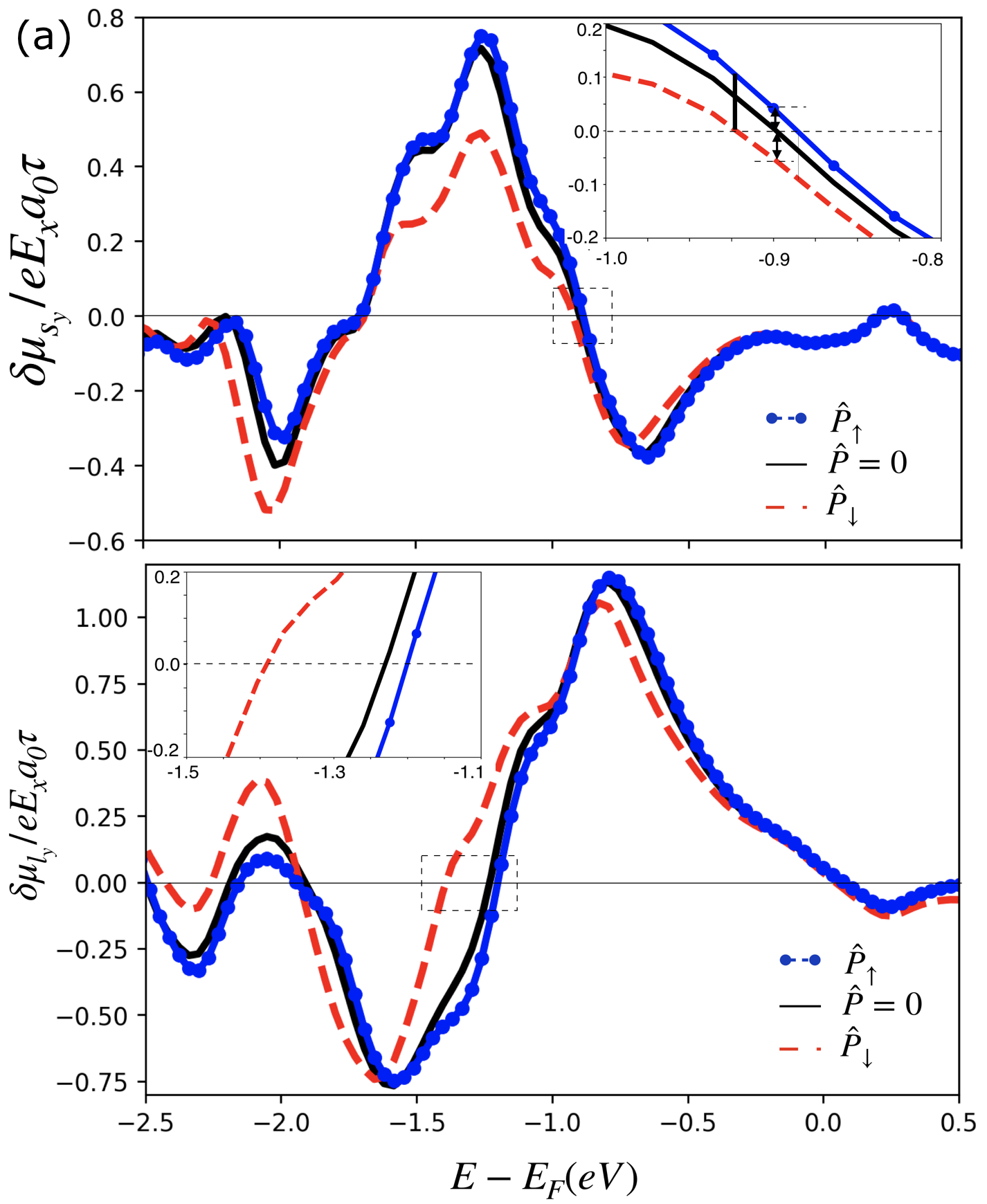}
    \caption{PtSe$_2$/MoSe$2$ heterostructure calculations of the (a) Spin density generated out-of-equilibrium using linear response. The inset shows the energy window at which we can see the sign change for different cases of the polarization field simulated by considering the potential slope previously extracted. (b) Orbital density generated out-of-equilibrium using linear response. The inset shows the energy window at which we can see the sign change for different cases of the polarization field simulated by considering the fit previously explained. The opposite sign displayed between spin and orbital degrees of freedom around [-1,0]~eV should be understood due to the negative spin-to-orbit correlation within the system.}
    \label{fig:SREE}
\end{figure}

We now turn to the case of PtSe$_2$/MoSe$_2$ structures deposited onto LiNbO$_3$. In a recent work~\cite{Massabeau2025}, we have performed the analysis of the spin dynamics in PtSe$_2$/MoSe$_2$ heterostructure, whereby using first-principles calculations alongside linear response theory, we have successfully explained the experimental THz-TDS measurements demonstrating an equal (in magnitude) and opposite electric field of approximately $0.2$ V/~\AA, enhancing the THz signal between the two different polarization directions. Furthermore, our results are in agreement with respect to the signal obtained experimentally in Co/Pt bilayers structures, where the primary mechanism for SCC is the inverse spin Hall effect. In the present case, our simulations shown in Fig.~\ref{fig:SREE}(a) and Fig.~\ref{fig:SREE}(b) reveal values for both spin and orbital accumulations respectively, which are on the same order of magnitude given the significant spin-orbit coupling of Pt. 

\noindent Additionally, on the spin side, we observed a considerable large energy window (approximately [-1.75, -1.0]~eV) for which the response with polarization P$_{\uparrow}$ directed towards the PtSe$_2$/MoSe$_2$ interface results in a larger THz signal compared to opposite polarization, labeled as P$_{\downarrow}$. At first glance, our estimation for this enhancement aligns with experimental results, indicating nearly a 30$\%$ increase in the emitted THz electric field amplitude. Furthermore, since our simulations are able to sample the behavior of the field-driven effect in a large energy window below the Fermi level, we identify an energy value of approximately 0.9~eV where the accumulation changes sign, meaning that at this energy level, it may be possible to reverse the sign of the THz signal by incorporating the polarization field from the ferroelectric material. Notably, we observed nearly opposite values of the spin density within the specified energy window, which could potentially be probed by tuning the photon energy used to generate the spin current on the ferromagnetic side, or even by varying the chemical potential by gating the system. 

\begin{figure}[!ht]
    \centering    \includegraphics[width=1\linewidth]{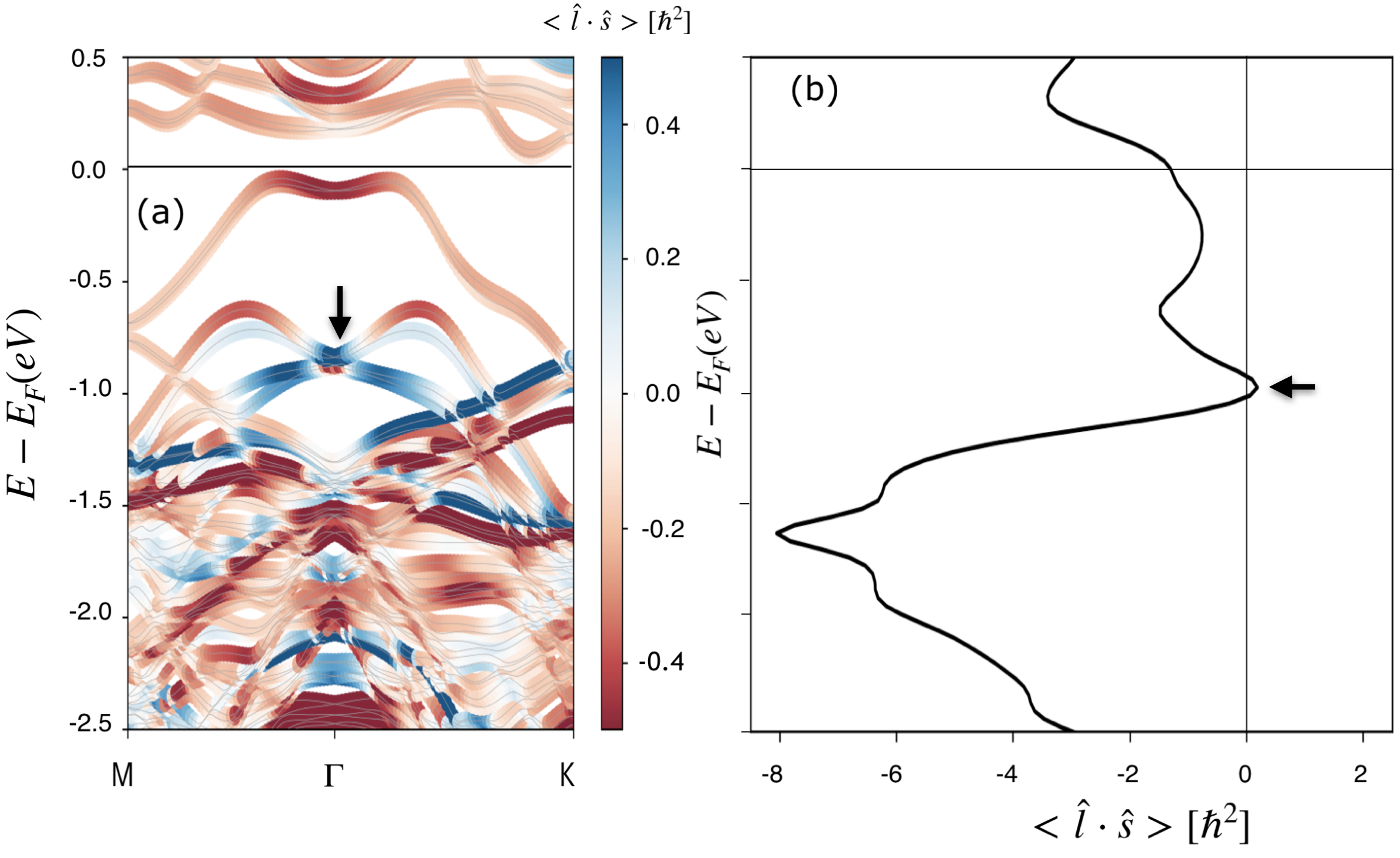}
    \caption{(a) Electronic band structure of the PtSe$_2$/MoSe$_2$ bilayer where the colormap represents the spin-to-orbit correlation along the high-symmetry path $M-\Gamma-K$, besides the (b) integrated value of the same quantity through the Brillouin zone. The negative value is in agreement with the opposite signals displayed by the spin and orbital accumulations. The black arrows signalized the energy value where a positive correlation around gamma compensates the negative behavior leading to a slightly net positive spin-to-orbit correlation displayed in (b).}
    \label{fig:soc_correlation}
\end{figure}

Similarly, the orbital density shown in Fig.~\ref{fig:SREE}(b) indicates an opposite value of the orbital-to-charge conversion compared to its spin counterpart. This difference is qualitatively explained by the negative spin-to-orbit correlation ($\braket{\hat{l}\cdot \hat{s}}$) illustrated in Fig.~\ref{fig:soc_correlation}, where the bands and the spin-to-orbit correlations are plotted side by side.  In Fig.~\ref{fig:soc_correlation} (a), the projection of $\braket{\psi_{\mathbf{k}n}|\hat{l}\cdot \hat{s}|\psi_{\mathbf{k}n}}$ correlation onto the electronic bands is shown, limited to the chosen energy window for out-of-equilibrium spin and orbital densities calculated along a high symmetry line like $M-\Gamma-K$. In Fig.~\ref{fig:soc_correlation} (b), we present the energy dependence of $\int_{n\mathbf{k}} \partial_{\varepsilon}f \braket{\psi_{\mathbf{k}n}|\hat{l}\cdot \hat{s}|\psi_{\mathbf{k}n}}$ spin-to-orbit correlation as a function of the Fermi level. From Fig.~\ref{fig:soc_correlation} (b) we can infer that the opposite sign between spin and orbital responses in the respective Fig.~\ref{fig:SREE}(a) and Fig.~\ref{fig:SREE}(b) corresponds to the negative correlation that persists along the chosen energy window.\\

\noindent By examining the out-of-equilibrium responses, we observe that the integration of the ferroelectric material modifies both spin and orbital responses in the 2D heterostructure. Notably, this modulation appears to be slightly more pronounced for the orbital degree of freedom, primarily due to changes in the charge distribution at the interface and the subsequent modifications once the ferroelectric material is considered. Specifically, the insets shown in Fig.~\ref{fig:SREE} highlight the energy values at which both responses change sign. The inset in Fig.~\ref{fig:SREE}(a) indicates the zero-accumulation crossing at 0.9~eV, where small displacements in energy between the three different cases, namely with and without the polarization field, are observed. In contrast, a more significant shift is found for the orbital response, as depicted in the inset in Fig.~\ref{fig:SREE}(b), occurring at around -1.5~eV. Moreover, the magnitude of the spin accumulation as a function of energy tends to increase more substantially between different field directions compared to the orbital counterpart at the same energy level. 
Additionally, our calculations suggest that it is possible to tune the sign of the spin-to-charge conversion, achieving nearly equal and opposite values for different directions of the polarization field. This characteristic would offer an interesting method for controlling spin transport in these structures.

\vspace{0.1in}

Regarding the strength of spin-to-charge conversion from IREE in these systems, it was reported in a previous study of PtSe$_2$/Gr heterostructures that the band splitting in the valence band achieves values comparable to a Rashba coupling of approximately $\alpha_R \approx -195$~meV$\cdot Å$. Although Pt possesses a larger spin-orbit coupling strength, this scenario also applies here, where a rough evaluation yields $\alpha_R \approx -300$~meV$\cdot Å$. This estimation primarily arises from the Se-$p$ states, as Se atoms are located at the interface, making their bands more sensitive to the interfacial dipole.\\


The magnitude of the emitted THz electric field due to spin (angular momentum)-to-charge conversion , $\mathbf{\mathcal{E}}$ is scaled by the corresponding inverse Rashba-Edelstein lengths, either $\Lambda^{s}_{REE}$ and $\Lambda^{l}_{REE}$ for the respective spin and orbital contributions according to the following relationships:


\begin{equation}§
    \hat{\mathcal{E}}\propto \Lambda^{s}_{REE}\left(\hat{\mathbf{\mu}}_s \times \hat{M}/|\hat{M}|\right),
    \label{Eq:spin_REE}
\end{equation}

\begin{equation}
    \hat{\mathcal{E}}\propto \Lambda^{l}_{REE}\left(\hat{\mu}_l \times \hat{M}/|\hat{M}|\right).
    \label{Eq:orbital_REE}
\end{equation}

\begin{figure}[!ht]
    \centering    \includegraphics[width=\linewidth]{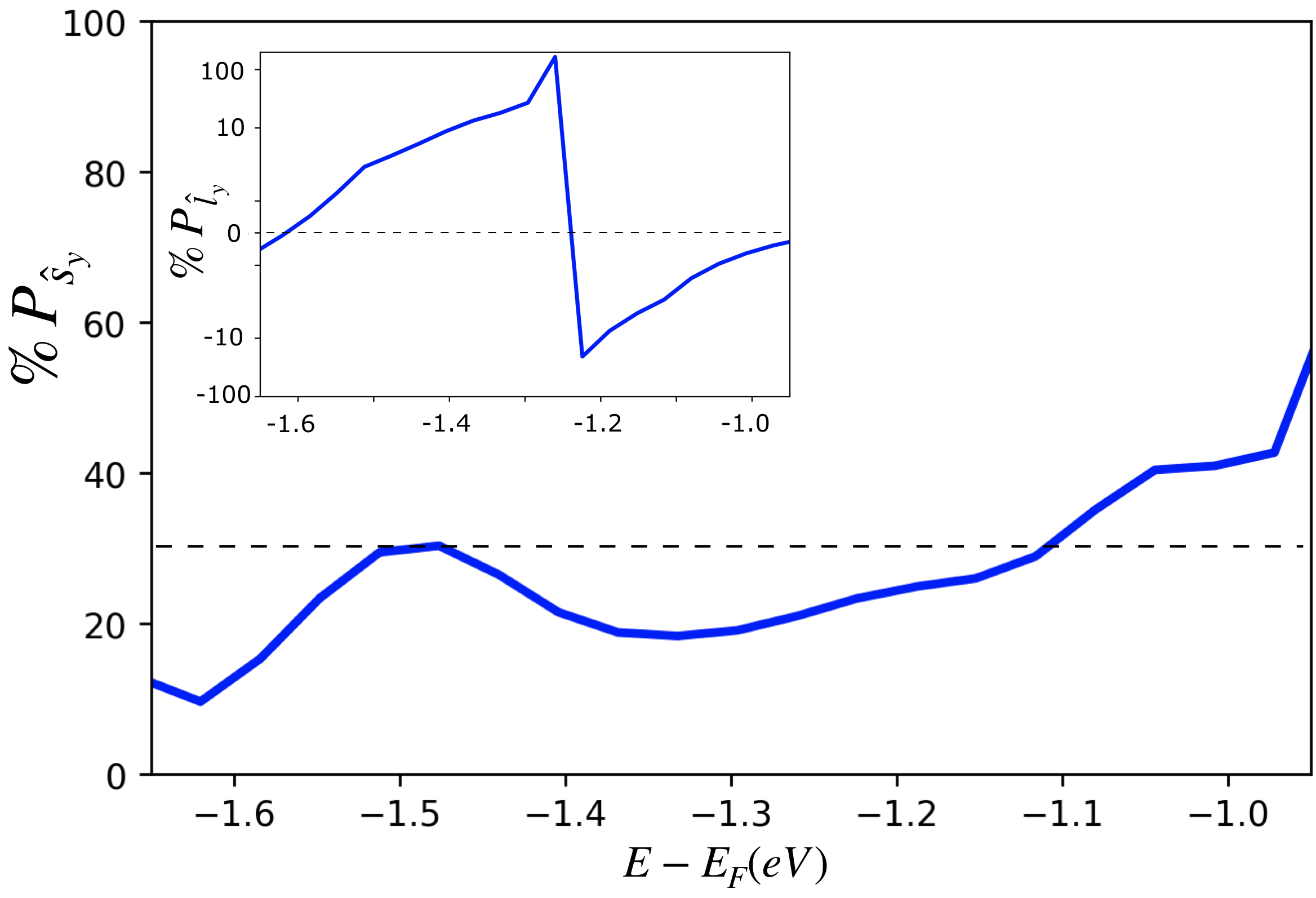}
    \caption{
    Percentage gain between different polarizations, that is, $\%$ $P_{\hat{s}_y}$ for the spin accumulation displaying a positive gain through the considered energy window $[-1.65,-0.95]$ eV, mainly lying below 30$\%$ as indicated by the dashed line. The inset shows the orbital counterpart for which we observe a larger positive $\% P_{\hat{l}_y}$ below $-1.2$~eV and a sign change ($\% P_{\hat{l}_y}<0$ ) for energy larger than $-1.2$~eV.}
    \label{fig:porcentage}
\end{figure}

The amplitude of the response is thus related to the respective spin ($\hat{\mathbf{\mu}}_{s_y}$) and orbital ($\hat{\mu}_{l_y}$) components of the angular momentum accumulation generated by the ultrashort laser excitation, then relaxing onto the polarized Fermi surface as obtained previously from the band structure calculations. In this sense, by considering Eq. \ref{Eq:spin_REE} and Eq. \ref{Eq:orbital_REE} we note that each of the responses is weighted by the amount of spin and orbital densities. We have considered equivalent characteristic relaxation times for the spin and orbit onto the Fermi surface of the Rashba-split 2D materials. Our simulations for the electric field-driven effect display a picture consistent with the behavior of both spin and orbital textures. This is depicted in Fig.~\ref{fig:porcentage} where we show the gain for the inverse spin Rashba-Edlestein length $\Lambda_{REE}^{s}\left(\hat{P}_{\uparrow}\right)$ with respect to $\Lambda_{REE}^{s}\left(\hat{P}_{\downarrow}\right)$, namely when the polarization field is either $\hat{P}_{\uparrow}$ or $\hat{P}_{\downarrow}$ respectively. The quantity $\%P$ is then defined as $\% P^s = \left(\frac{\Lambda_{REE}^{s}\left(\hat{P}_{\uparrow}\right)-\Lambda_{REE}^{s}\left(\hat{P}_{\downarrow}\right)}{\Lambda_{REE}^{s}\left(\hat{P}_{\uparrow}\right)+\Lambda_{REE}^{s}\left(\hat{P}_{\downarrow}\right)}\right)\times 100\%$. Overall we observe a positive $\%$P lying below $\%30$ with a sign changing around a narrow energy window at the vicinity of $-0.9$~eV. On the other hand, the orbital polarization counterpart $\% P^l = \left(\frac{\Lambda_{REE}^{l}\left(\hat{P}_{\uparrow}\right)-\Lambda_{REE}^{l}\left(\hat{P}_{\downarrow}\right)}{\Lambda_{REE}^{l}\left(\hat{P}_{\uparrow}\right)+\Lambda_{REE}^{l}\left(\hat{P}_{\downarrow}\right)}\right)\times 100\%$, defined in the same manner, is shown in the inset where the signal change occurs below $-1.2$~eV, but again displays a predominantly positive value for $\%P$ in the energy window where the experimental SCC has to be considered. Similarly, the orbital part displays a more susceptible response with respect to the polarization field, reaching an enhancement of $\sim$ $\%$ 100 demonstrating a more sensitive response compared to the spin. In general, we notice that within the energy window that spans in the $[-1.65,-0.95]$~eV range, there is an enhancement in the case of the positive field in agreement with the experiment.

\vspace{0.1in}

So far, in our analysis of PtSe$_2$-based heterostructures, we have found that incorporating the orbital density generated out of equilibrium leads to a response that aligns more closely with experimental observations from THz-TDS measurements. Although spin-orbit correlation indicates opposing behaviors within specific energy ranges, the two responses balance each other out, resulting in a net positive accumulation of both spin and orbital densities. Nevertheless, at this stage, if the spin-polarization generated by the optical pulse is pretty well-known considering the magnetization/polarization of the ferromagnet, it seems difficult to precisely determine the subsequent orbital polarization generated likewise, because involving the spin-orbit strength of the different constituting materials (in the bulk or at the interface with the ferromagnet).

That is, although the electrical driven response is on the same order of magnitude as its spin counterpart, what is usually imprinted from the magnetization dynamics, corresponds to an orbital injection of an order of magnitude smaller compared to the spin which forbids the clean summation of both responses. However, this can still be tailored in order to maximize the orbital contribution by considering elements such as Ni, which was demonstrated to be a good orbital momentum injector able to perform orbital pumping effects in light metal interfaces \cite{hayashi2023observation}.

\section{Conclusions}

In summary, we have investigated the spin and orbital transport properties at the interfaces of PtSe$_2$/MoSe$_2$, specifically when influenced by the incorporation of ferroelectric materials like LiNbO$_3$ on the vdW heterostructure interface. By employing, we estimated the effective electric field generated by the presence of the ferroelectric material. Given that ferroelectric domains have been observed experimentally, we examined both interfaces to account for the polarization field in both directions. This approach provides valuable insights into the electronic properties of these heterostructures, allowing us to calculate the out-of-equilibrium spin and orbital accumulation densities resulting from the inversion symmetry breaking at the interface and the polarization field. Based on our simulations, we discovered a significant energy window in which the spin-charge conversion mechanism explains recent experimental results obtained using THz time-domain spectroscopy. Our findings quantitatively match the observed effects of the substrate polarization fields as well as the same conversion sign in both PtSe$_2$/Gr and PtSe$_2$/MoSe$_2$ interfaces.

\

Additionally, we identified that the response signal could be tuned based on the energy of the incident photons used to generate hot holes from the ferromagnetic side. We compared the generated densities using our approach for the PtSe$_2$/Graphene interface, which yielded the same results previously obtained \cite{Abdukayumov2024}. We believe this method can be adapted to systems and can be further utilized to propose material engineering strategies for manipulating spin and orbital transport in vdW heterostructures. We have also addressed and quantified the possible role played by the orbital degree of freedom on the THz response finding orbital-to-charge conversion efficiency comparable to that of the spin contribution, with however the actual limitation for the quantification of the exact amount of the orbital polarization generated by the optical pulse. In order to achieve a better understanding of experimental outcomes, it becomes essential to obtain more information about the orbital relaxation time and the degree of orbital injection from the FM side. Since this property is material-dependent, we believe it should be possible to engineer spintronic samples where the orbital response is predominant. In this context, we have outlined a methodology for understanding the spin and orbital-to-charge conversion mechanisms. Finally, we consider the implications of taking into account the dynamics of both degrees of freedom in the context of the direct electrically driven effect, specifically the SOT. Recent studies \cite{Canonico2025} indicate that by lifting the spin degeneracy in centrosymmetric PtSe$_2$, we can achieve significant efficiencies compared to other 2D materials. This advancement may also influence the research on other transition metal dichalcogenides, such as PtTe$_2$~\cite{Yadav2024}.

\acknowledgments
 
We thank L. Vojáček for fruitful discussions. This work was supported by the ANR 'ORION' project, grant ANR-20-CE30-0022-01 of the French Agence Nationale de la Recherche, ANR SR-'Equipex+2D-MAG' project, grant ANR-21-ESRE-0025, ANR-22-CE30-0026 'DYNTOP' project, by a France 2030 government grant managed by the French National Research Agency PEPR SPIN ANR-22-EXSP0009 'SPINTHEORY', ANR-22-EXSP-0003 'TOAST' and ANR-22-EXSP0011 'OXIMOR', and by the EIC Pathfinder OPEN grant 101129641 'OBELIX'. Part of the calculations were performed using computational resources provided by GENCI–IDRIS (Grants 2025-A0170912036 and  2025-AD010916348).

\bibliography{refs-resub-2,bibliography}
\end{document}